\documentstyle[12pt]{article}
\begin{document}
\input{epsf}
\begin{titlepage}
\begin{flushright}
DAMTP 96-04 \\
KL-TH 3/96
\end{flushright}
\vspace{3ex}
\begin{center}
{\LARGE \bf
\centerline{The Roughening Transition of the 3D Ising }
\centerline{ Interface: A Monte Carlo Study}
}
\vspace{0.65 cm}
{\large     M. Hasenbusch, }

\vspace{0.17 cm}
{\it  DAMTP, Silver Street, Cambridge, CB3 9EW, England}

\vspace{0.65 cm}
{\large  S. Meyer and M. P\"utz    }

\vspace{0.17 cm}
{\it Fachbereich Physik, Universit\"at Kaiserslautern ,  Germany}

\end{center}
\setcounter{page}{0}
\thispagestyle{empty}
\begin{abstract}\normalsize
We study the roughening transition of an interface in an Ising
system on a 3D simple cubic lattice using a finite size scaling method. 
 The particular method
has recently been proposed and successfully tested for various solid on solid 
models. The basic idea is the matching of the renormalization-group-flow
of the interface  with that of the exactly
solvable body centered cubic solid on solid model. 
We  unambiguously
confirm the Kosterlitz-Thouless nature
of the roughening transition of the Ising interface.
Our result for the inverse transition temperature $K_R=0.40754(5)$ 
is almost by two orders of magnitude more accurate than
the estimate of Mon, Landau and Stauffer \cite{mon90a}.
\end{abstract}
\nopagebreak
\vspace{1ex}
\end{titlepage}

\newpage
                       
\section{Introduction} 
\addcontentsline{toc}{chapter}{Introduction}
  Interfaces play an essential role in many areas of science. In condensed
  matter physics, the most prominent examples are surfaces of liquids 
  and solids.
  In 1951 Burton, Cabrera and Frank \cite{burton51a} pointed out that a 
  phase transition may occur in the equilibrium structure of crystal 
  surfaces.     
  Such a phase transition from a smooth to a rough surface is called 
  a roughening transition. 
 They viewed a growing layer of a crystal as a two dimensional Ising model. 
 The part of the layer  occupied by atoms  is represented by spin +1,
 while the vacancies are represented by  spin -1.  
 From the exact solution of the two-dimensional 
 Ising model \cite{onsager} one then infers 
the existence of a phase transition. 
 However this picture of a crystal surface is very crude. Obviously in a real
 crystal surface there are more than just one single incomplete layer of
 atoms.   

 A better description of a crystal in equilibrium with its 
 vapour is provided by the three-dimensional Ising model, where spin
 +1 represents a site occupied by an atom, while spin -1 represents a 
 vacancy. The boundary conditions are chosen such that an interface 
 between a region with most spins equal to +1 and a region with most spins 
 equal to -1 is present. 
 In 1973 Weeks et  al. \cite{weeks73a} performed a low temperature 
 expansion                                           
 for the width of an (001) interface  in a three-dimensional Ising 
 model on a simple cubic lattice with isotropic couplings.
 They obtained a roughening temperature $T_R = 0.57 \;\; T_c$,
 where $T_c$ is the temperature of the bulk phase transition,
 while the approximation of the interface by 
 the two dimensional Ising model yields $T_R = 0.503  \;\; T_c$. 
The best known value for the inverse of
the critical temperature of the three-dimensional Ising model on
a simple cubic lattice is given by $\beta=0.2216546(10)$ \cite{blote}.

 A fairly good approximation of the Ising interface is given by the 
 so called SOS (solid on solid) models. Neglecting overhangs of the 
 interface and bubbles in the bulk, the variables of the two 
 dimensional models give the height of the interface measured above some
 reference plane.  
 A duality transformation exactly relates these models with two
 dimensional $XY$ models \cite{savit}. In contrast to the two-dimensional
 Ising model which undergoes a second-order phase transition, 
  SOS models, as $XY$ models,  are 
 expected to undergo a 
 Kosterlitz-Touless (KT) phase transition \cite{KT}. 

%

 The important question is whether overhangs of the interface and 
bubbles in the bulk-phases are irrelevant for the critical behaviour 
of the interface in the Ising model and hence the roughening transition 
is of KT nature as it is the case for SOS models. 
 Monte Carlo simulations of the interface        
 in a three-dimensional Ising model have been performed in order 
 to answer this question
  \cite{kner83a,mon88a,mon90a}.

However, 
the numerical determination of the transition temperature and the 
confirmation of the KT nature of the phase transition has turned out to be
extremely difficult.  The reason for this problem can be found in the 
KT theory itself. At the roughening  transition corrections  are present
that vanish
logarithmically with the length scale. Therefore for all lattices sizes
that can be simulated one is quite far away from the Gaussian fixed point
that is the basis for the formulas used for fits in refs. 
\cite{kner83a,mon88a,mon90a}. 
In order to overcome this problem one has to take into account the logarithmic
corrections in the analysis of the data \cite{around}.

However one can do even better.  The body centered cubic
solid on solid (BCSOS) model that was introduced by van Beijeren in 1977
\cite{beijeren77} as a solid on solid approximation of an interface in
an Ising
model on a body centered cubic lattice on a (001) lattice plane is solved 
exactly.
The BCSOS model is equivalent with one of the exactly solved vertex models
\cite{lieb,wu,baxter}. The exact formula for the free energy and the correlation
length proves the model to have a KT-type phase transition. 

In \cite{physica,thesis} it was proposed to match the renormalization group
(RG)-flow  of other SOS
models and the Ising interface with that of the BCSOS model. The RG-flow is
monitored by properly chosen block-observables on finite lattices. In order
to determine the roughening temperature of an SOS model
 or the Ising interface
one has to find the temperature
where the values for the block-observables for the BCSOS model at 
the roughening temperature are reproduced.

This method has been successfully applied to the ASOS (absolute value SOS)
model, the discrete Gaussian model and the dual of the standard $XY$ model in 
2 dimensions \cite{physica}. Preliminary results were also obtained for the 
Ising interface \cite{thesis}. 

In the present paper we improve the result of ref. \cite{thesis} by 
drastically
increasing the statistics for the Ising model as well as for the BCSOS model. 
The largest lattice size considered is increased from $64 \times 64 \times 27$
 to
 $256 \times 256 \times 31$ 
allowing the direct matching of the Ising interface with the BCSOS model
compared with the indirect approach via the ASOS model of
ref. \cite{thesis}. 
The high statistical accuracy needed for the matching method could be obtained
in a moderate amount of CPU time (about two months on a workstation) due to the 
use of highly efficient cluster algorithms for the Ising interface \cite{prl}
as well as for the BCSOS model \cite{loop}.
 
This paper is organized as follows. In section 2
 we define the models to be studied. 
We summarize  exact results for the BCSOS model \cite{lieb,wu,baxter}
relevant to our study.
We briefly discuss the RG-flow diagram  of the KT phase-transition. 
In section 3 we describe the matching method. 
We give 
particular emphasis to the special problems arising  in the case of
the Ising interface. In section 4 we discuss  
our numerical results.
A comparison with previous Monte Carlo studies is presented in section 5.
In section 6 we give our conclusions and an outlook. 

\section{The models}
 We consider an Ising model on a simple cubic lattice
 with extension $L$ in x- and
 y-direction and with extension $D$ in z-direction.
 For reasons given by the algorithm 
 we shall only consider odd values of $D$.
 We have chosen the convention that 
 the z-coordinate of the lattice points takes the 
 half-integer values $z=-D/2 \; ... \; D/2$. 
The Ising model is defined by the partition function
\begin{equation}
Z=\sum_{s_i=\pm 1} \exp(-K^I H) \; ,
\end{equation}
where the classical Hamiltonian is given by
\begin{equation}
  H = - \sum_{<ij>} J_{<ij>} s_i s_j \; ,
\end{equation}
where the summation is taken  over 
all nearest neighbour pairs $<ij>$ on the lattice, and $K_I = \frac{1}{k_BT}$
is the normalized inverse temperature. 
 In x- and y-direction we consider periodic boundary conditions.
 In order to create an interface we apply anti-periodic boundary conditions
 in the remaining z-direction. 
Anti-periodic boundary conditions are defined by $J_{<ij>}=-1$ for bonds $<ij>$
connecting the lowermost and  uppermost plane of the
lattice, while all other nearest neighbour pairs keep $J_{<ij>} = 1$. 



The ASOS model is the solid on solid approximation of an interface
in an Ising system on a simple cubic lattice on a (001) lattice plane.
It is defined by the partition function
\begin{equation}
 Z = \sum_{h_i} \exp(-K^{ASOS} \sum_{<i,j>} |h_i - h_j|) \; ,
\end{equation}
where  $h_i$ is integer-valued and the summation is taken over the nearest
neighbour pairs of a 2D square lattice. 
At  low temperatures the Ising interface and the ASOS model are related by
$2 K^I =  K^{ASOS}$.

In the case of the BCSOS model the 
 two-dimensional lattice splits in two sub-lattices like a
checker board. In the original formulation, on one of the sub-lattices the
spins takes integer values, whereas the spins on the other sublattice
take half-integer values. We adopt a different convention: spins
on ``odd'' lattice sites take values of the form $2n+(1/2)$, and spins
on ``even'' sites are of the form $2n-(1/2)$, $n$ integer.
As a consequence,
the effective distribution for block spins (= averages over blocks)
will be centered around integer values (instead of half integer values),
and the average of the lowest energy configurations take
 integer values like it is the
case for the ASOS models defined above.
The partition function of the BCSOS model can be expressed as
\begin{equation}
  Z = \sum_{h} \exp(-K^{S} \sum_{[i,k]} |h_i - h_k|)  \; ,
\end{equation}
where $i$ and $k$ are next to nearest neighbours. Nearest
neighbour spins $h_i$ and $h_j$ obey the constraint
$|h_i - h_j| = 1$.
Van Beijeren \cite{beijeren77} has shown
that 
the configurations of the BCSOS model are in one-to-one correspondence
to the configurations of the F model, which is a special six vertex model.
The F model can be solved exactly
with transfer matrix methods \cite{lieb,wu,baxter}.
For our choice of the field variable 
 the roughening coupling is given by
\begin{displaymath}
 K_R^{S} = \frac12 \ln2 .             
\end{displaymath}
The critical
behaviour of non-local quantities such as the
correlation length $\xi$ is known and has the form predicted by KT
theory \cite{baxter}.

 

 
 The RG-flow of SOS models is well described by two     
 parameters  $\beta$ and $y$ \cite{KT}.
 The two dimensional Sine Gordon model is especially suited to
 discuss the flow of these parameters with the length scale, since
 this model contains $\beta$ and $y$ as bare parameters in its action.
 On the lattice it is given by the partition function 
\begin{equation}
 Z^{SG} = \int [\mbox{d} \phi] \;
    \exp \left(-\frac 1 {2 \beta} \sum_{<i,j>} (\phi_i - \phi_j)^2
         + y \sum_i \cos (2 \pi \phi_i) \right), 
\end{equation}
 where the $\phi_i$ are real numbers.
  
For the continuum version of the model with a momentum cutoff
one can derive the 
 parameter flow under infinitesimal RG transformations \cite{KT}.    
 It is given, to second order in perturbation theory, by 
\begin{equation}
 \dot{x}  = - z^2 \; , \;\;\;\;\;\;\;\;\;\;\;\; \dot{z} = - xz \;,
\end{equation}
 where  $z  = const \cdot y$  and $x  = \pi \beta - 2$. 
 $const$ depends on the particular cutoff scheme used. 
 The derivative is taken with respect to the logarithm of the 
 cutoff scale. 
For large $x$,  $z$ flows towards $z=0$. The large distance behaviour of 
the model is therefore the same as that of the massless Gaussian model.
For small $x$,  $z$ increases with
increasing length scale. The theory is therefore massive.  The critical 
trajectory separates these two regions in the coupling space. It ends at a
Gaussian fixed point  characterized by $x=0$ or $\beta=\frac{2}{\pi}$. On 
the critical trajectory the fugacity vanishes as
\begin{equation}
\label{fug}
 z(t) = \frac{1}{z_0^{-1} + t} \; , 
\end{equation}
where $t$ is the logarithm of the cutoff scale. 
 
\section{The matching method}\label{uni}

In order to compare the RG-flows of the Ising interface
and the BCSOS model we follow closely the method introduced in ref.
 \cite{physica}.
 This method is closely related 
 to the finite size scaling 
 methods proposed by Nightingale \cite{nightingale} and Binder \cite{binder}. 
 No attempt is made to compute the RG-flow of the couplings explicitly,
 but rather the RG-flow is monitored by evaluating quantities 
 that are primarily 
 sensitive to the lowest frequency fluctuations on a finite lattice. 

In order to separate the low frequency modes of the field
a block-spin transformation \cite{kadanoff} is used.
Blocked systems of size $l \times l$ are considered. The size $B$ of a
block (measured in units of the original lattice spacing)
is then given by
$B = L/l$, where $L$ is the linear size of the original lattice.
For solid on solid models the block-spins are defined by
\begin{equation}
\phi_{\tilde j} = B^{-2} \sum_{j \in \tilde j} h_j  \;, 
\end{equation}
where $\tilde j$ labels square blocks of a linear extension $B$. 
One should note that this linear blocking rule has the half-group property
that the successive applications of two transformations  with a scale-factor
of $B$  have exactly the same effect as a single transformation  with a 
scale-factor of $B^2$.  
 
  Since the position of the interface in the Ising system
 is not well defined on a
  microscopic level we have to look for a substitute of the blocking
  rule applied to the field of the SOS models. In the following we briefly
  discuss the solution of the problem proposed and applied to the study
 of interfaces 
 in the rough phase in ref. \cite{tension}.  

 First we  have to  ensure that the interface is not located at 
 the $z=-D/2$ to $z=D/2$ boundary, in which case the definition of the
  blockspin  discussed 
 below would become meaningless. 
 Therefore we locate the interface in the system in
 a crude fashion, which is done by searching for the z-slice with the 
 smallest absolute value of the magnetisation. Then we redefine the $z$-
 coordinates such  that the interface is located close to $z=0$.
 Now one can go ahead with the measurement of interface properties ignoring 
 the periodicity of the lattice in z-direction. 

 The blocks considered  have the full lattice extension D in z-direction
 and an extension $B$ in x- and y-direction.
 The  interface position is defined inside a block by
 \begin{equation}
   \phi_{\tilde i}= \frac{M_{\tilde i}}{2 m B^2} \; ,
   \label{define}
 \end{equation}
 where $M_{\tilde i}$ is the total magnetisation in the block $\tilde i$
and $m$ is the bulk magnetisation.
This definition is motivated by the naive picture that
above the interface the magnetisation takes  uniformly the value $-m$ and
below the value $m$.

One has to discuss to what extent the meaning of the above definition is
spoiled by corrections to this simple picture. What are the effects of bulk
fluctuations and overhangs of the interface?
The fluctuations of the bulk magnetisation are given by the magnetic
susceptibility $\chi$. The square of the fluctuations of $M_{\tilde i}$
 induced by
the bulk fluctuation is therefore given by
$ \sigma^2(M_{\tilde i}) = D B^2 \chi$
and the resulting fluctuations of $\phi_{\tilde i}$ induced by bulk fluctuations
is given by
\begin{equation}
\sigma^2_b (\phi_{\tilde i}) =   \frac14  \frac{D}{B^2}  \frac{\chi}{m^2} \; .
\end{equation}
As a consequence the larger the block size in x- and y-direction, 
the better the position
of the interface is defined. However, when $D$ is sent to infinity for a fixed
$B$, the position of the interface defined by eq. (\ref{define}) 
becomes meaningless.

In ref. \cite{prl,physica} we proposed a radical solution to overcome the
problem of bulk fluctuations completely.  Before the position of the
interface is determined all bubbles are removed. Technically this is
accomplished  by performing standard cluster-updates at $K=\infty$.
Since the absolute value of 
the bulk magnetisation becomes 1 in this process the definition of
the interface position is modified to
 \begin{equation}
   \bar{\phi}_{\tilde i}= \frac{\bar{M}_{\tilde i}}{2 B^2} \; .
 \end{equation}
Overhangs of the interface are expected on a scale of the bulk correlation
length $\xi_b$. Therefore we have to assume that the position of the
interface only has a well defined meaning if the blocksize is large compared
with the bulk correlation length $ B >> \xi_b $.  At the roughening transition
this is however no severe restriction since $\xi_b \approx 0.3 $.

Finally one should ensure that no additional interfaces are 
created spontaneously. Following ref. \cite{compa} the surface tension 
at $K^I=0.40$ 
is $\sigma=0.6721(1)$.
This is certainly a lower bound for the  surface tension at the roughening
coupling to be found below. 
  Therefore two additional 
interfaces are suppressed by at least a factor of 
$ D^2 \exp(-0.6721 \times 2L^2)$, which is certainly sufficient already 
for the smallest lattice size $L=32$ that we consider. 

Now we define suitable observables for the blocked systems discussed above. 
Motivated by the perturbation theory of the Sine-Gordon model
two types of observables are chosen:
such that are ``sensitive'' to the flow of the kinetic
term (flow of $K$), and such that are sensitive to the
fugacity (periodic perturbation of a massless Gaussian
model). For the first type of observables we choose
\begin{equation}\label{a1}
A_1 = \langle  (\phi_{\tilde i}-\phi_{\tilde j})^2 \rangle \;, 
\end{equation}
where $\tilde i$ and $\tilde j$ are nearest neighbours on the block lattice, 
and
\begin{equation}\label{a2}
A_2 = \langle  (\phi_{\tilde i}-\phi_{\tilde k})^2 \rangle \; ,
\end{equation}
where $\tilde i$ and $\tilde k$ are next to nearest neighbours.
Note that these quantities are only defined for $l > 1$.
As a monitor for the fugacity we take the following set
of quantities (defined for $l=1,2,4$):
\begin{eqnarray}\label{a3}
  A_3 &=& \langle  \cos (2\pi \phi_{\tilde i}) \rangle
  \nonumber \\
  A_4 &=& \langle  \cos (2 \cdot 2\pi \phi_{\tilde i}) \rangle
  \nonumber \\
  A_5 &=& \langle  \cos (3 \cdot 2\pi \phi_{\tilde i}) \rangle
\end{eqnarray}
\subsection{Determination of the roughening coupling}
As discussed in ref. \cite{physica} there are
 two parameters which have to be adjusted in order
to match the RG flow of the Ising interface with that of the
critical BCSOS model:
The  coupling $K^I$ of the Ising interface
and  in addition
the ratio $b=B^{I}/B^{S}=L^{I}/L^{S}$ of the lattice sizes (and hence the
 block sizes) of
the Ising model and the BCSOS model.
A $b \neq 1$ is
needed to compensate for the
different starting points of the two models on the critical RG-trajectory.
In ref. \cite{physica,thesis} $b = 3.3(3)$ was obtained. 
 
For the proper values of the roughening coupling $K^I_R$ and
the matching constant $b$ observables of the Ising interface and the 
BCSOS model match like
\begin{equation}
A_{i,l}^{I}(b \; B,K^{I}_R) =
A_{i,l}^{S}(B,K^{S}_R)  + O(B^{-\omega}) \;, 
\end{equation}
where $i$ indicates
the observable and $l$ 
the size of the blocked lattice. 
The $O(B^{-\omega})$ corrections are due to irrelevant operators.
$\omega$ is the correction to scaling exponent. 
The perturbation theory of the Sine-Gordon model suggests $\omega=2$. 

In order to obtain a numerical estimate for the roughening coupling $K^{I}_R$
of the Ising model and the matching factor $b$ for a given lattice size $L^{S}$
of the BCSOS model we required that the equation above is exactly fulfilled
for two block observables. 

In the following we considered the pairs 
$(A_{1,l},A_{3,l})$ and $(A_{2,l},A_{3,l})$ for  $l=2$ and $l=4$. 
Replacing $A_{3,l}$ by $A_{4,l}$ or $A_{5,l}$ leads to statistically 
poorer results. 

We solved the system of two equations for the two observables
$A_{i,l}$ and $A_{j,l}$ numerically by first computing the $K^I_{i,l}(b)$
and $K^I_{j,l}(b)$ 
that solve the single equations for a given value of $b$. The intersection 
of the two curves $K^I_{i,l}(b)$ and $K^I_{j,l}(b)$ gives us then the
solution of the system of two equations.  For an illustration of this method
see figures 5 and 6 of ref. \cite{physica}. 

In \cite{physica} it was demonstrated, that 
the corrections to scaling for the observables
$A_1$ and $A_2$ for SOS models are similar to
those in the massless continuous Gaussian model. Therefore
we considered  the  ``improved'' observable $D_1$
which is defined as follows:
\begin{equation}
D_1(L) = \frac{A_1^{(0)}(\infty)}{A_1^{(0)}(L)} A_1(L)
\end{equation}
$A_1^{(0)}$ is computed for the massless Gaussian model defined by
\begin{equation}
Z_0 = \int \prod_x d \psi_x \exp( - \frac12 \sum_{<x,y>} (\psi_x-\psi_y)^2 )
\end{equation}
An improved quantity $D_2$ is defined analogously.
 Explicit results for  $A_1^{(0)}$               
 and $A_2^{(0)}$ are given in table 4 of ref. \cite{physica}. 
   
%
Obviously this modification does not affect the large $L$ behaviour
since $A_{1}^{(0)}(L)=A_{1}^{(0)}(\infty)+O(L^{-2})$. In the following 
we will see that the results for our largest lattice sizes are virtually 
unaffected by this kind of improvement. However the benefit
for the smaller lattices will be clearly visible.

\section{Discussion of the numerical results}
First we simulated the BCSOS model at the roughening coupling
$K_R = \frac12 \ln 2$
on square lattices of the size $L \times L$
with periodic boundary conditions imposed. 
We considered lattices with sizes ranging from
$L=8$ up to $L=96$. 
The loop algorithm of Evertz, Marcu and Lana
\cite{loop}  
enabled us to reduce the statistical error of the BCSOS data 
compared with those given in \cite{physica} by a factor of about 5. 
We performed $4 \cdot 10^6$ measurements throughout, except for $L=96$, where 
$3 \cdot 10^6$ measurements were performed.  The number of loop-updates
per measurement was 10 for $L=8$ up to 35 for $L=96$.  This number was chosen 
such that the integrated autocorrelation times in units of measurements
were about 1. These simulations took about 81  hours of CPU time on a
IBM RISC System/6000 Modell 590 (66 Mhz) in total. 
The results for $l=4$ for the critical BCSOS model are presented in figures
2 to 6. 
The numbers for the $A$'s can be obtained from the authors on request.

Next we performed the simulations of the Ising model
at the best known value $K_R=0.4074$
\cite{thesis} for the roughening coupling. We used the modified 
cluster algorithm introduced in \cite{prl}. For a discussion of the 
algorithm see also ref. \cite{tension}.
The expectation values for $K^I$ in the neighbourhood of the simulation
point are then  obtained by
re-weighting \cite{swendsen-ferrenberg}.

First of all we had to ensure that our results are not spoiled by 
a to small extension in z-direction. 
Obviously the thickness of the lattice has to be large compared to 
the width of the interface itself. It turns out that this requirement
can be easily fulfilled since the width of the interface at $K_R$ is 
smaller than 1 for the lattice size we will consider in the 
following \cite{kner83a,prl,mon90a,tension}.  

In \cite{prl,tension,compa} the dependence of interface properties  on 
the extension of the lattice in z-direction  was carefully checked. 
For couplings close to $K_R$ and lattice sizes $L \le 256$ 
it turned out that
for about $D > 10$ no dependence on $D$  can be detected within the 
obtained accuracy. 

We performed an additional check for $L=128$. We compared the interface
width without bubbles
for $D=15$ and $D=31$. We obtain  $W_0^2=0.64647(25)$ for $D=15$ 
and $W_0^2=0.64658(33)$ for $D=31$.
The definition of $W_0^2$ is given below.

Therefore we regard the extension $D=31$ which we used throughout
in the following simulations as perfectly safe.

Per measurement we performed 5 H-updates and 5 I-updates and in 
addition one Metropolis sweep.  Remember that H and I refer
 to reflections at
half-integer and integer z-values respectively. 
The number of updates per sweep was again 
chosen such that the integrated autocorrelation times 
in units of measurements 
are smaller than 1. The number of measurements was 100000 for most of the 
lattice sizes. For $L=48$, $192$ and $L=256$ we performed $ 93500$, 
$113000$ and $72500$ measurements respectively. 
The total amount of CPU time used for the Ising simulations 
was about 63 days on an IBM RISC System/6000 Modell 590 (66 Mhz). 

\subsection{The interface width}
Before discussing the results of the matching analysis let us briefly  
study the behaviour of the surface width, which was the basis of the 
Monte Carlo study of ref. \cite{mon90a}.
Following ref. \cite{mon90a} a normalized magnetization gradient is 
defined as
\begin{equation}
\rho(z)= \frac{1}{2 m_b L^2} [M(z+1/2) - M(z-1/2)] \;  , 
\end{equation}
where $m_b$ is the bulk magnetization and  $M$ the total 
magnetization of a z-slize of the lattice. 
Now the interface width is given by 
\begin{equation}
 W^2 = \langle \; \sum_z \rho(z) z^2 -
 \left( \sum_z \rho(z) \rho(z) z  \right)^2\;\rangle\;.
\end{equation} 
We computed the interface width for the original configurations ($W^2$) and 
after removing the bubbles ($W_0^2$). 
In the rough phase of the Ising interface or an SOS model  the surface 
width behaves like
\begin{equation}
\label{width}
 W^2 = const + \frac{1}{2 \pi \beta_{eff}} \ln L
\end{equation} 
for sufficiently large $L$. $\beta_{eff}$ is determined by the point where 
the RG-trajectory hits the $x$ axis in the $KT$ flow diagram.  Therefore 
the slope of the curve at the roughening transition should be given by
 $1/\pi^2$.
However it was pointed out in refs. \cite{physica,around}
that at the roughening transition corrections to  this asymptotic behaviour 
die out proportional to the fugacity  (\ref{fug}) i.e. only proportional
to the inverse of the logarithm of $L$.
Hence these corrections cannot be neglected even for quite large 
lattice sizes.

\begin{figure}[]

\epsfxsize=14cm
\centerline{\epsfbox{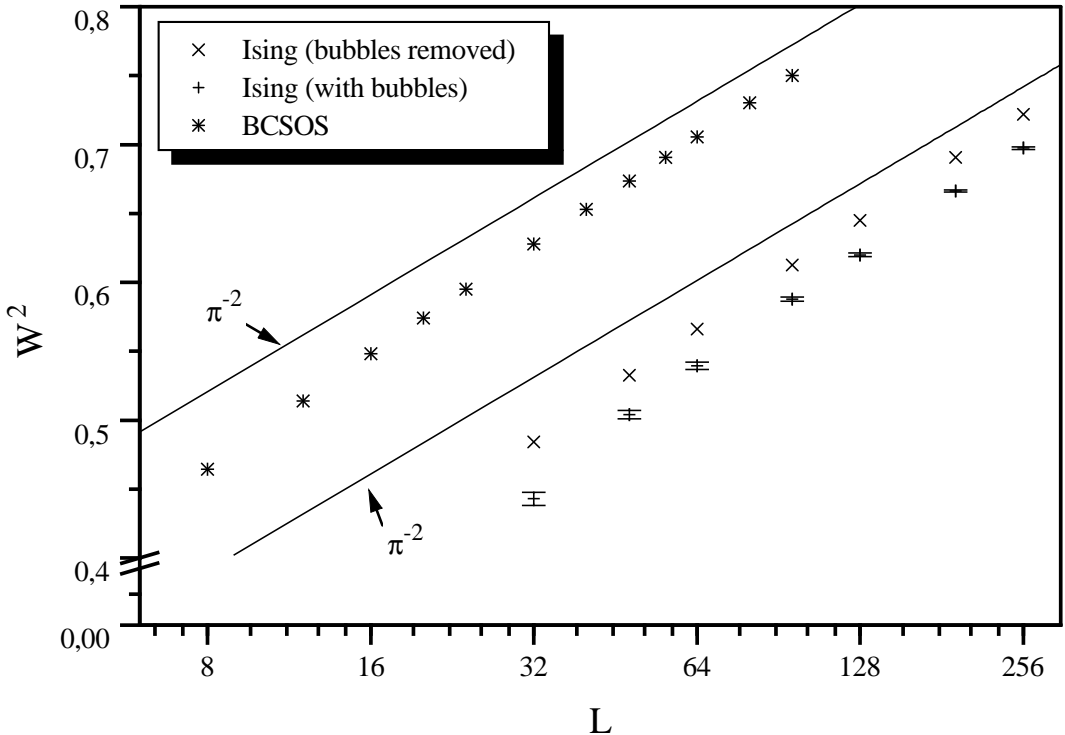}}
\caption{\label{schichtdicke}
 { \sl Squared interface width $W^2$ plotted versus the lattice size $L$ }
}
\end{figure}

 To check this anticipated behaviour we have plotted the 
surface thickness for the BCSOS model and the Ising interface with bubbles and 
bubbles removed in figure \ref{schichtdicke}. For comparison two straight lines
with slope $1/\pi^2$ are given.   It turns out that even when our largest 
lattice sizes are considered the  slope is by about $10\%$ larger than the 
asymptotic one. Using therefore eq. (\ref{width}) with the slope $1/\pi^2$
as criterion to determine the roughening  point leads to a considerable 
under-estimation of the roughening temperature.  Going to enormous lattice 
sizes such as $L=960$ does not help that much to overcome the problem since the 
corrections die out only logarithmically in  $L$. 

\subsection{Matching results}
For the determination of the roughening temperature and
the matching factor using the matching method we have 
considered the four pairs of observables $(A_{1,2}, A_{3,2})$, 
$(A_{2,2}, A_{3,2})$, $(A_{1,4},  A_{3,4})$ and $(A_{2,4},  A_{3,4})$. 
In order to check the effect of removing the  bubbles and the improvement
by the continuous Gaussian model on our results
we performed the whole analysis for the following
four different choices of the observables: \\
\phantom{ii}i)
 The bubbles being removed from the bulk phases,  using the improved
$D_1$ and $D_2$.\\
\phantom{i}ii) The bubbles
 being removed from the bulk phases,  using  $A_1$ and $A_2$.\\
iii) Not removing the bubbles, 
using the improved $D_1$ and $D_2$. \\
iv) Not removing the bubbles, using  $A_1$ and $A_2$.

In all cases the
 statistical errors of the estimates for the roughening coupling and 
the matching factor were computed from a Jackknife analysis put on top
of the whole matching procedure. 

The results for the roughening coupling $K_R$ for the
Ising interface  based on i) and ii)
are summarized in table \ref{krittbl}. 
Using the improved quantities $D_1$ and $D_2$ the results 
are consistent within two standard deviations starting from $L=96$ for 
$l=2$ as well as $l=4$. 
 For our largest lattice size $L=256$ the difference
of the results when using the unimproved observables $A_1$ and  $A_2$ 
rather than $D_1$ and $D_2$ is smaller than the statistical
error. 
However looking at the results obtained from $L=32$ and $L=48$ 
makes clear that a major part of the corrections to scaling is eliminated
by using $D_1$ and $D_2$ instead of $A_1$ and  $A_2$. 

\begin{table} \label{krittbl}
 \caption{Results for the roughening coupling obtained from matching
for different values of $L$ and $l=2,4$.
For the matching $A_{3,l}$ together 
with either $A_{1,l}$ , $D_{1,l}$ , $A_{2,l}$ or  $D_{2,l}$
is used. The bubbles in the Ising system are removed before the measurement.}
 \begin{center}
  \begin{tabular}{|c|c|c|c|c|c|}
   \hline
   $L$ & $l$ & $A_1$ & $D_1$ & $A_2$ & $D_2$ \\
   \hline
    32 & 2 &   0.40602(15) &   0.40727(14) &   0.40657(16) &   0.40831(21) \\
    48 & 2 &   0.40648(16) &   0.40716(16) &   0.40678(17) &   0.40760(21) \\
    64 & 2 &   0.40675(14) &   0.40722(14) &   0.40692(15) &   0.40735(18) \\
    96 & 2 &   0.40719(14) &   0.40741(15) &   0.40727(16) &   0.40745(17) \\
   128 & 2 &   0.40733(15) &   0.40748(16) &   0.40734(19) &   0.40748(21) \\
   192 & 2 &   0.40762(14) &   0.40768(14) &   0.40762(15) &   0.40766(16) \\
   256 & 2 &   0.40748(14) &   0.40751(14) &   0.40762(16) &   0.40765(16) \\
   \hline
    32 & 4 &   0.40529(10) &   0.40678(9)  &   0.40589(10) &   0.40740(10) \\
    48 & 4 &   0.40610(9)  &   0.40709(8)  &   0.40647(9)  &   0.40748(9) \\
    64 & 4 &   0.40649(8)  &   0.40728(8)  &   0.40677(9)  &   0.40750(9) \\
    96 & 4 &   0.40702(8)  &   0.40745(8)  &   0.40717(8)  &   0.40754(9) \\
   128 & 4 &   0.40727(7)  &   0.40752(7)  &   0.40733(8)  &   0.40755(9) \\
   192 & 4 &   0.40753(7)  &   0.40766(7)  &   0.40754(7)  &   0.40764(8) \\
   256 & 4 &   0.40738(7)  &   0.40745(7)  &   0.40741(7)  &   0.40746(7) \\
   \hline
  \end{tabular}
 \end{center}
\end{table}

In table \ref{matchtbl} we present 
the results for the matching factor $b$ based on i) and ii). 
The observations  are analogous to those for the roughening 
coupling.

\begin{table} \label{matchtbl}
 \caption{Results for matching parameter $b$ 
for different values of $(L)$ and $(l=2,4)$. 
For the matching $A_{3,l}$ together
with either $A_{1,l}$ , $D_{1,l}$ , $A_{2,l}$ or  $D_{2,l}$
is used.  Bubbles are removed from the Ising configurations.}
 \begin{center}
  \begin{tabular}{|c|c|c|c|c|c|}
   \hline
   $L$ & $l$ & $A_1$ & $D_1$ & $A_2$ & $D_2$ \\
   \hline
    32 & 2 & 2.12(4)  & 2.68(5)  & 2.346(6) & 3.14(11) \\
    48 & 2 & 2.30(8)  & 2.74(12) & 2.49(10) & 3.05(16) \\
    64 & 2 & 2.54(11) & 2.92(12) & 2.68(11) & 3.02(16) \\
    96 & 2 & 2.60(14) & 2.84(18) & 2.69(17) & 2.88(20) \\
   128 & 2 & 2.85(19) & 3.05(23) & 2.86(24) & 3.05(29) \\
   192 & 2 & 2.74(24) & 2.82(27) & 2.73(26) & 2.80(28) \\
   256 & 2 & 3.08(33) & 3.15(35) & 3.45(44) & 3.54(48) \\
   \hline
    32 & 4 & 1.751(10) & 2.048(11) & 1.864(13) & 2.194(18) \\
    48 & 4 & 2.062(18) & 2.419(22) & 2.191(24) & 2.548(28) \\
    64 & 4 & 2.311(26) & 2.703(32) & 2.442(33) & 2.816(42) \\
    96 & 4 & 2.550(41) & 2.87(6)   & 2.66(5)   & 2.95(7) \\
   128 & 4 & 2.68(6)   & 2.91(7)   & 2.73(7)   & 2.94(8) \\
   192 & 4 & 2.81(9)   & 2.96(10)  & 2.82(10)  & 2.94(11) \\
   256 & 4 & 3.05(11)  & 3.16(12)  & 3.10(13)  & 3.19(13) \\
   \hline
  \end{tabular}
 \end{center}
\end{table}

For a block-size $B \ge 32$ the results obtained from iii) and iv)
are consitent within error-bars with those obtained from i) and ii) .
We therefore skip a detailed discussion of these results.

Following the above discussion we regard the results for $K_R$ and $b$ 
obtained from i) as the least affected by systematical errors. 
Therefore we take the weighted average of the results of i) 
(i.e. the $D_1$ and $D_2$ columns of table 1 and 2) for $l=2$, 
$L\ge 96$ and $l=4$, $L \ge 128$ as our final result. 

We obtain $K_R = 0.40754(5)$ and $b=2.97(7)$.  Computing the error-estimate
we have assumed that values obtained from the same  simulation
but from a different pair of observables are strongly correlated.

\begin{figure}[]

\epsfxsize=14cm
\centerline{\epsfbox{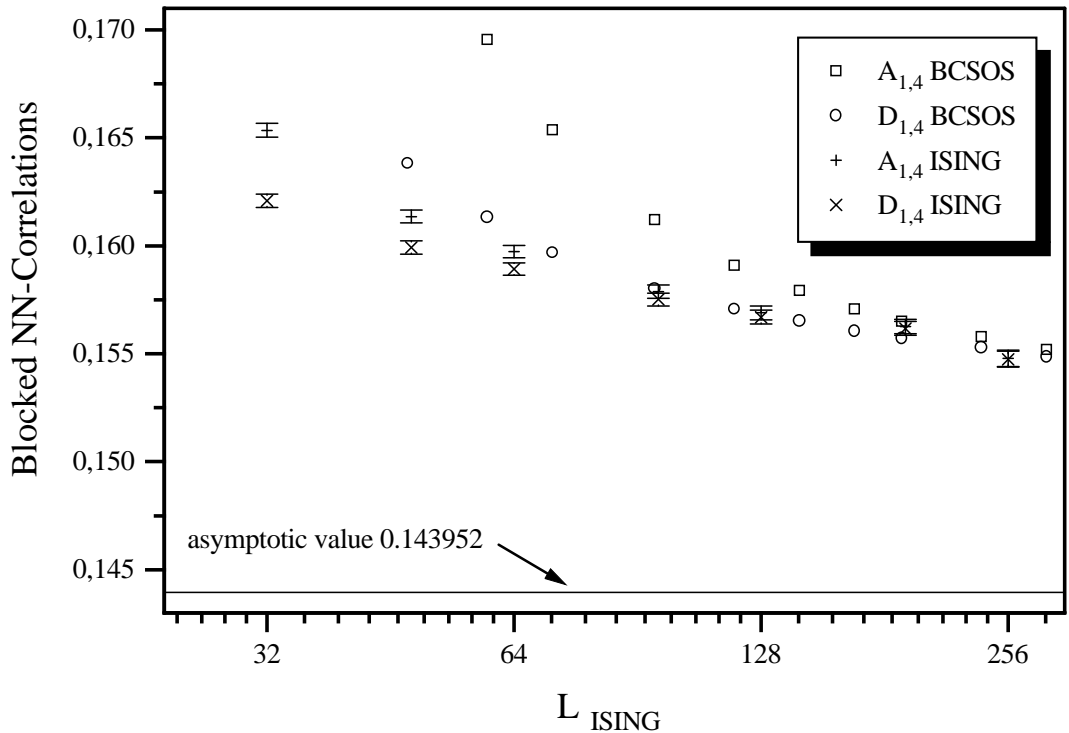}}
\caption{\label{matchA14D14}
 { \sl $A_{1,4}$ and $D_{1,4}$ at the roughening transition as a function 
 of $L^{I} = b~L^{S}$ for the Ising interface as well as the BCSOS
 model }
}

\end{figure}

As an additional check of universality we have plotted 
 all observables $A_{i}$ considered for the 
critical BCSOS model as well as the Ising interface  at $K_R = 0.40754(5)$
as a function of $L$ and $L/b$ respectively 
in figures \ref{matchA14D14}-\ref{matchA54}.

In figure \ref{matchA14D14} the observables $A_{14}$ and $D_{14}$ are plotted.
The figure shows nicely the improvement gained by replacing 
$A_{14}$ by $D_{14}$. While the curves of $A_{14}$  for the Ising interface 
and the BCSOS model are only consistent for $L\le 192$, 
the consistency extends down to at least $L=96$ when 
$D_{14}$ is considered. For $A_{24}$ and $D_{24}$ given in figure
\ref{matchA24D24}  this improvement is visible even more drastically. 

In the case of 
$A_{34}$ plotted in figure \ref{matchA34}
the matching of the curves within the statistical accuracy 
sets in for $L^I\ge 64$. 

In figure \ref{matchA44} the observable $A_{44}$ is plotted. This observable
was not used for the determination of $K_R$ and $b$.  The fact that
for $L^I \ge 96$ the curves of $A_{4,4}$ for the Ising interface and the 
BCSOS model fall on top of each other, strongly supports that the Ising 
interface and the BCSOS model have the same  critical behaviour. 
For $L^I \ge 128$  also the curves of $A_{5,4}$ for the Ising 
interface and the
BCSOS model become identical within error bars.

Similar observations hold for $l=1$ and $l=2$. (The correponding figures
are not reproduced here.) 

\begin{figure}[]

\epsfxsize=14cm
\centerline{\epsfbox{ad24.eps}}
\caption{\label{matchA24D24}
 { \sl $A_{2,4}$ and $D_{2,4}$  at the roughening transition as a function
 of $L^{I} = b~L^{S}$ for the Ising interface as well as the BCSOS
 model }
}

\end{figure}

\begin{figure}[]

\epsfxsize=14cm
\centerline{\epsfbox{a34.eps}}
\caption{\label{matchA34}
 { \sl $A_{3,4}$ at the roughening transition as a function
 of $L^{I} = b~L^{S}$ for the Ising interface as well as the BCSOS
 model }
}
\end{figure}

\begin{figure}[]

\epsfxsize=14cm
\centerline{\epsfbox{a44.eps}}
\caption{\label{matchA44}
 { \sl $A_{4,4}$ at the roughening transition as a function 
 of $L^{I} = b~L^{S}$ for the Ising interface as well as the BCSOS
 model }
}
\end{figure}

\begin{figure}[]

\epsfxsize=14cm
\centerline{\epsfbox{a54.eps}}
\caption{\label{matchA54}
 { \sl $A_{5,4}$ at the roughening transition as a function 
 of $L^{I} = b~L^{S}$ for the Ising interface as well as the BCSOS
 model }
}
\end{figure}

The matching programme also allows to determine the nonuniversal constants
appearing in formulas describing the divergence of observables
near the roughening transition. 
The critical behaviour of the correlation length $\xi$
is described by \cite{KT}
 \begin{equation}\label{xxii}
     \xi = A \; \exp( C \; \kappa^{-1/2}) \;\;.
 \end{equation}
Along the lines of section 5.3 of ref. \cite{physica}
we obtain $A^{I} = 0.74(2)$ and 
$C^{I} =  1.03(2)$ for the Ising interface.

\section{Comparison with existing studies}
Let us compare our result, $K_R=0.40754(5)$, with results
obtained in previous Monte Carlo studies of the Ising interface.

  We nicely confirm the value $K_R=0.4074(3)$ obtained by one of the 
  present authors \cite{thesis} using similar 
  methods as used in the present paper.   

  In a large scale Monte
  Carlo simulation of lattices up to $960 \times 960
  \times 26$   Mon, Landau and Stauffer \cite{mon90a} studied
  the behaviour of the interface width. 
  They obtain $K_R^I = 0.409(4)$.

  Mon, Wansleben, Landau and Binder  \cite{mon88a} determined the  step
  free energy on lattices of size up to $96^3$. They give the estimate 
  $T_R/T_c=0.54(2)$ which corresponds to $K_R = 0.410(16)$. 

  B\"urkner and Stauffer \cite{kner83a} obtained in their pioneering study 
  $T_R/T_c=0.56(3)$ which corresponds to $K_R = 0.396(22)$. 
 
  One should note that all these estimates are consistent within the quoted
  error-bars with our present result. 

  It is also interesting to compare the Ising interface results with those 
  obtained in ref. \cite{physica} for the ASOS model, 
  $K_R^{ASOS}= 0.8061(3)$   and $b_m^{ASOS} = 2.8(3)$.  In the low temperature
  limit the ASOS coupling and that of the Ising model are related as
  $K^I=1/2 K_{ASOS}$.  At finite temperatures one expects that the bubbles
  and the overhangs disorder the Ising interface compared to the ASOS model.
  Hence one expects the roughening transition of the Ising interface
  to occur at a lower temperature as for the ASOS model and indeed
  $K_R^I - 1/2 K_R^{ASOS} = 0.0044(2)$.  

  Such a shift was already
  predicted by low temperature expansions \cite{adler}, $K_{R,LT}^I =
  0.404(12)$ and  $K_{R,LT}^{ASOS} = 0.787(24)$.
  However one should note that the difference of these two results 
  is  smaller than the given error-bars. 
  The matching factors for the two models 
are equal within the quoted error-bars.  The direct matching 
of the Ising interface with the ASOS model performed in ref. \cite{thesis} 
gives the more precise result $b_m^I/b_m^{ASOS} = 1.17(4)$  
for this particular comparison. 

\section{Conclusion and outlook} 
Using the finite size scaling method of ref. \cite{physica} we
unambigously demonstrated the KT-nature of the roughening transition in
 the (001) interface of the 3D Ising model.  Our estimate for the inverse 
of the roughening temperature $K_R = 0.40754(5)$ is almost by  a factor of 100
more accurate then the best previously published value \cite{mon90a}. 

Previous studies have been mainly plagued by logarithmic corrections at the 
roughening transition. 
This problem  has been overcome completely. 

In addition the use of efficient, virtually 
slowing-down free, cluster algorithms for the BCSOS model \cite{loop} and
the Ising interface \cite{prl,tension} allowed us to generate more than $10^6$
independent configurations in the case of the BCSOS model and about $10^5$ 
independent configurations  for the Ising model for all lattice sizes 
using about 2 month of CPU time on an IBM RISC System/6000 Modell 590 (66 Mhz). 

This high statistical accuracy can be further improved. First simply  by
using more CPU time and secondly by further improvements in the implementation 
of the algorithm.  


\section*{Acknowledgements}

M. Hasenbusch thanks M. Marcu and K. Pinn for sharing their insight 
in Wilsons RG with him.
M. Hasenbusch expresses his gratitude
for support by the Leverhulme Trust under grant  
16634-AOZ-R8 and by PPARC.

Most of the simulations were performed on a IBM RS6000 cluster of the 
RHRZ (Regionales Hochschul-Rechenzentrum  Kaiserslautern).

\newpage
{\large \bf Figure Captions}

\vspace{0.5 cm}
  {\bf Fig.1} \\
  { Squared interface width $W^2$ plotted versus the lattice size $L$ }

\vspace{0.5 cm}
  {\bf Fig.2} \\
 {  $A_{1,4}$ and $D_{1,4}$ at the roughening transition as a function 
 of $L^{I} = b~L^{S}$ for the Ising interface as well as the BCSOS
 model }

\vspace{0.5 cm}
  {\bf Fig.3} \\
 {  $A_{2,4}$ and $D_{2,4}$  at the roughening transition as a function
 of $L^{I} = b~L^{S}$ for the Ising interface as well as the BCSOS
 model }

\vspace{0.5 cm}
  {\bf Fig.4} \\
 {  $A_{3,4}$ at the roughening transition as a function
 of $L^{I} = b~L^{S}$ for the Ising interface as well as the BCSOS
 model }

\vspace{0.5 cm}
  {\bf Fig.5} \\
 {  $A_{4,4}$ at the roughening transition as a function 
 of $L^{I} = b~L^{S}$ for the Ising interface as well as the BCSOS
 model }

\vspace{0.5 cm}
  {\bf Fig.6} \\
 {  $A_{5,4}$ at the roughening transition as a function 
 of $L^{I} = b~L^{S}$ for the Ising interface as well as the BCSOS
 model }

\end{document}